\documentclass[12pt]{ronbun}

\usepackage{amsmath,amssymb}
\usepackage{cite}
\usepackage{bm}
\usepackage{dcolumn}

\newcommand{\nn}{\nonumber}

\newcommand{\del}{\delta}

\newcommand{\al}{\alpha}

\renewcommand{\th}{\theta}

\newcommand{\s}{\scriptscriptstyle}

\newcommand{\thetab}{{\bar\theta}}

\numberwithin{equation}{section}

\begin{document}

\begin{flushright}
\parbox{4.2cm}
{KEK-TH-895 \hfill \\
{\tt hep-th/0306213}
 }
\end{flushright}

\vspace*{1.1cm}

\begin{center}
 \Large\bf Dirichlet Branes of the   
Covariant Open Supermembrane \\
on a PP-wave Background
\end{center}
\vspace*{1.5cm}
\centerline{\large Makoto Sakaguchi$^{a}$ and Kentaroh Yoshida$^{b}$}

\begin{center}
\emph{Theory Division, High Energy Accelerator Research 
Organization (KEK),\\
Tsukuba, Ibaraki 305-0801, Japan.} 
\\
\vspace*{1cm}
$^{a}$Makoto.Sakaguchi@kek.jp
~~~~
 $^{b}$kyoshida@post.kek.jp
\end{center}

\vspace*{2cm}

\centerline{\bf Abstract}
  
\vspace*{0.5cm}
We discuss an open supermembrane theory on the maximally supersymmetric
pp-wave background in eleven dimensions. The boundary surfaces of an
open supermembrane are studied by using the covariant supermembrane
theory. 
In particular, we find the configurations of M5-branes and
9-branes preserving a half of supersymmetries at the origin.

\vfill
\noindent {\bf Keywords:}~~{\footnotesize open supermembrane,
Wess-Zumino term, pp-wave}

\thispagestyle{empty}
\setcounter{page}{0}

\newpage 

\section{Introduction}

Supermembrane theory \cite{BST,dWHN} is expected to play a fundamental
role in M-theory.  The Banks-Fischler-Shenker-Susskind matrix model
proposed in \cite{BFSS} is also related to the supermembrane theory
through the matrix regularization.  We can generalize the matrix model
to that on the general consistent background that has non-vanishing
curvature and includes non-trivial fluxes.  However, it is too difficult
to study it due to the complicated interaction terms.  In the recent
progress, the matrix model on the maximally supersymmetric pp-wave
background was proposed by Berenstein, Maldacena and Nastase \cite{BMN},
and has been analyzed extensively by various authors \cite{DSR,SY1,alg}.

Hitherto, {\it closed} supermembranes on the pp-wave background were
widely studied from the various viewpoints in the works
\cite{DSR,SY1,SY2}. On the other hand, {\it open} supermembranes have
not been understood well yet.  In the work \cite{SY1} boundary
conditions of an open supermembrane were studied and the possible
Dirichlet branes were classified.  As the result, 1/2 BPS Dirichlet
$p$-branes are allowed for the value $p=1$ only. It should be remarked
that this result holds for the Dirichlet branes sitting {\it outside}
the origin.  If we consider branes sitting {\it at} the origin, then we
can find the configurations of M5-branes and 9-branes preserving a half
of supersymmetries, as we will see later. In fact, the D-branes sitting
at the origin in superstring theories on pp-waves are special ones among
other D-brane configurations as noted by several authors
\cite{BP,BGG,GG,BPZ,SMT}.  This situation comes from the fact that the
homogeneity of pp-waves is not manifest in Brinkmann coordinates used in
the analysis.  Because pp-waves are homogeneous, branes sitting at the
origin can be located at any arbitrary other location.  But this is
manifest only in Rosen coordinates.  In particular, a brane sitting at
(outside) the origin in Brinkmann coordinates is mapped to a flat
(curved) brane in Rosen coordinates.

It was difficult to possess an overall picture of D-brane configurations
on pp-waves since the analysis are based on the light-cone gauge fixed
Green-Schwarz superstring action.  However, the authors of the work
\cite{BPZ} proposed the method to covariantly analyze D-branes on
pp-waves (following Ref.\,\cite{LW}).  Motivated by this work, we
analyze covariantly the Dirichlet branes of an open supermembrane on the
pp-wave background.  The covariant representation of the Wess-Zumino
term is constructed up to and including the fourth order in the
$SO(10,1)$ spinor $\th$, and then we study the boundary conditions for
Dirichlet branes.  In this formalism we can of course rederive the
well-known result in flat space: Dirichlet $p$-branes are allowed for
the values $p=1,5$ and $9$ only \cite{EMM,dWPP}.  The value $p=5$ means
that open M2-branes can end on the M5-brane, and the value $p=9$ implies
the end of the world 9-brane in the Horava-Witten theory \cite{HW}.  The
physical meaning of $p=1$ case has not been well understood.  In the
case of pp-waves, the analysis gets complicated because of surface terms
which originate from the non-vanishing curvature and constant flux. In
this paper, we find the configurations of M5-branes and 9-branes
preserving a half of supersymmetries {\it at} the origin. These are not
1/2 BPS objects outside the origin as noted in Ref.\,\cite{SY1}.  The
9-branes sitting at the origin support the work \cite{M} about the
heterotic matrix model.

This paper is organized as follows.  In section 2 we introduce the
covariant expression of the Wess-Zumino term in the supermembrane theory
on the pp-wave background.  In section 3 we will study the boundary
surface (Dirichlet brane) of an open supermembrane and classify the
possible configuration of them. In particular, M5-branes and 9-branes
are 1/2 BPS objects at the origin (while those are not 1/2 BPS outside
the origin).  The 1-branes preserve a half of supersymmetries at the
origin and even outside the origin.  Section 4 is devoted to a conclusion
and discussions.  In Appendix our notation and convention are summarized.

\section{Covariant Action of Supermembrane on the PP-wave} 

In this section,
we briefly review the relevant aspects of the supermembrane theory on the
maximally supersymmetric pp-wave background.
We will obtain the covariant representation of the Wess-Zumino term
up to and including fourth order in the $SO(10,1)$ spinor $\th$,
which will be analyzed in the next section. 

The Lagrangian of the supermembrane \cite{BST,dWHN} is formally given as
a sum of the Nambu-Goto type Lagrangian $\mathcal{L}_0$ and Wess-Zumino
term $\mathcal{L}_{WZ}$\footnote{Our notation and convention are
summarized in Appendix.}
\begin{eqnarray}
&& \mathcal{L} \;=\;  \mathcal{L}_0
+ \mathcal{L}_{WZ}\,,\quad \mathcal{L}_0 \;=\;
- \sqrt{-g(X,\th)}  \,,
~~~
\mathcal{L}_{WZ} \;=\;\!B
\label{Lag}
\label{start}
\end{eqnarray}
where the induced metric $g_{ij}$ is the pull-back of the background
metric $G_{MN}$
\begin{equation}
 g_{ij} \; = \; E_i^M E^N_j G_{MN}
 =E_i^A E^B_j \eta_{AB}, \quad g \;=\; {\rm det}\,g_{ij}\,.
\end{equation}
and the three-form $B$ is defined by $H=dB$ with $H$ being the pull-back
of the four-form gauge superfield strength on the supergravity
background.  The term ${\cal L}_0$ is manifestly spacetime
superinvariant while ${\cal L}_{WZ}$ is quasi-superinvariant,
superinvariant up to a surface term.  The $\kappa$-invariance of the
action must be imposed in order to match the fermionic and bosonic
degrees of freedom on the world-volume.  It is known that the condition
for the $\kappa$-invariance of the action is equivalent to the
supergravity equation of motion.  For an open supermembrane, the
$\kappa$-variation leads to a surface term, which must be deleted by
imposing appropriate boundary conditions on the boundary of the open
supermembrane.

We shall consider supermembranes on the maximally supersymmetric pp-wave
solution of the eleven dimensional supergravity, called as the
Kowalski-Glikman (KG) solution \cite{KG},
\begin{eqnarray}
\label{KG}
&&ds^2 \,=\, 
- 2 dx^+ dx^- + G_{++}(dx^+)^2 + \sum_{\mu =1}^9(dx^{\mu})^2\, , \\
& & \qquad G_{++} \,\equiv\, - \left[\left(\frac{\mu}{3}\right)^2 
 (x_1^2 + x_2^2 + x_3^2) + \left(\frac{\mu}{6}\right)^2 (x_4^2 + \cdots + 
 x_9^2)\right]\, , \nn\\
&& F_{+123} \,=\, \mu\, \quad (\mu \neq 0)\,,
\label{flux}
\end{eqnarray}
where $F_{+123}$ is a constant four-form flux pointing $+,\,1,\,2,\,3$
directions.
It is known \cite{BFHP2} that the KG solution can be
derived from $AdS_4 \times S^7$ or $AdS_7 \times S^4$ by a Penrose limit
\cite{P}.
The supervielbeins of the supermembrane on the $AdS_7\times
S^4$ background are \cite{C,dWPPS}
\begin{eqnarray}
& &
E^A \,=\, dX^M 
\tilde{e}^A_M
- i \bar{\th} \Gamma^A
\left(\frac{2}{\mathcal{M}}{\rm sinh}\frac{\mathcal{M}}{2}\right)^2
D\th\,, \quad
E^{\bar{\alpha}} \,=\, \left(
\frac{{\rm sinh}\mathcal{M}}{\mathcal{M}} D\th \right)^{\bar{\al}}\,, \\ 
& &
i\mathcal{M}^2 \; = \; 2(T_A{}^{B_1\cdots B_4}\th ) F_{B_1\cdots B_4}
(\bar{\th}\Gamma^A) \nn \\
&& \hspace*{2cm} - \frac{1}{288}
(\Gamma_{A_1A_2}\th)[
\bar{\th}(\Gamma^{A_1A_2B_1\cdots B_4}F_{B_1\cdots B_4}
+ 24\Gamma_{B_1B_2} F^{A_1A_2B_1B_2} 
)] \nn\\ 
&& (D\th)^{\bar{\alpha}} 
\,\equiv\, d\th^{\bar{\alpha}} 
+ \tilde{e}^{A}
(T_A{}^{B_1\cdots B_4}\th)^{\bar{\alpha}}
 F_{B_1\cdots B_4} 
- \frac{1}{4}\tilde{\omega}^{A_1A_2}
(\Gamma_{A_1A_2}\th)^{\bar{\alpha}}\,, 
\nn\\
& & T_A{}^{B_1\cdots B_4} \; = \; \frac{1}{288}
(\Gamma_A{}^{B_1\cdots B_4} - 8 \delta_A^{[B_1} \Gamma^{B_2B_3B_4]} )\,, \quad \tilde{e}^A
\,=\, dX^M\tilde{e}^A_M\,,\nn
\end{eqnarray}
where $\tilde{e}^A_M$'s and $\tilde{\omega}^{AB}$ are, respectively,
vielbeins and spin connection of the $AdS_7\times S^4$ background.
Moreover, the third rank tensor $B$ of $AdS_7\times S^4$ is given by
\begin{eqnarray}
B &=& B^{(1)} + B^{(2)}\,, \\
B^{(1)} &=& \frac{1}{6}e^{A_1}\wedge e^{A_2} \wedge 
e^{A_3}\, C_{A_1A_2A_3}\,, \\
B^{(2)} &=&  i \! \int^1_0\!\!\! dt\, \bar{\th}\Gamma_{AB}E(x,t\th)\wedge 
E^A(x,t\th)\wedge E^B(x,t\th)\,,
\label{ads-WZ}
\end{eqnarray}
where $E^A(x,t\th)$ and $E^{\bar{\al}}(x,t\th)$ is obtained by the shift
$\th \rightarrow t\th$ in $E^A$ and $E^{\bar{\alpha}}$.

Now let us construct the covariant Wess-Zumino term in the case of the
KG solution.
The supervielbeins and the three-form $B$ for the KG
solution can be extracted from those for AdS$_{4/7}\times S^{7/4}$
\cite{HKS:11}.  In the Penrose limit, the vielbeins
$\tilde{e}^{{A}}_{{M}}$ and the spin connection $\tilde \omega^{AB}$ of
$AdS_7\times S^4$ reduce to $e^{A}_{M}$ and $\omega^{AB}$ of the KG
solution
\begin{eqnarray}
 & & e^{A}_{M}: \quad\, e^{+}_{+} = e^{-}_{-} = 1, \quad e^{+}_{-} = 0,
 \quad e^{-}_{+} = - \frac{1}{2}G_{++}, \quad e^r_{\mu} =
 \del^r_{\mu}\,, \\ &&\omega^{AB}:~~ \omega^{r-}=\frac{1}{2}\partial^r
 G_{++}dX^+,~~~ \mbox{otherwise}=0.
\end{eqnarray}
The second term $B^{(2)}$ of the Wess-Zumino term (\ref{ads-WZ}) becomes
\begin{eqnarray}
B^{(2)}&=& \frac{i}{2}\thetab\Gamma_{AB}D\theta \wedge dX^Me_M^A
\wedge dX^Ne_N^B
+\frac{1}{2}\thetab\Gamma_{AB}D\theta \wedge
\thetab\Gamma^AD\theta \wedge dX^Me_M^B \nonumber\\&&
+\frac{i}{24}\thetab\Gamma_{AB}{\cal M}^2D\theta \wedge dX^Me_M^A
\wedge dX^Ne_N^B +{\cal O}(\theta^6).
\end{eqnarray}
We do not need to expand the first term $B^{(1)}$ in the Wess-Zumino term 
and the kinetic term ${\cal L}_0$ with respect to $\theta$,
because these terms turn out to be irrelevant in our analysis.
The covariant Wess-Zumino term constructed above will be used in the
next section for the classification of boundary surfaces of an open
supermembrane on the pp-wave background.

\section{Classification of supersymmetric Dirichlet Branes}

Here we will classify possible supersymmetric Dirichlet branes on the
pp-wave background by using the covariant open supermembrane.  In
conclusion, we will find supersymmetric M5-branes and 9-brane sitting at
the origin as Dirichlet branes of an open supermembrane on the pp-wave
background.  In addition, 1-branes are also allowed to exist and they
also preserve a half of supersymmetries at the origin. 

\subsection{Boundary Conditions of Open Supermembrane}

To begin with, let us introduce the boundary condition on the boundary
$\partial\Sigma$ of the world-volume $\Sigma$ of an open supermembrane.
The coordinates should be distinguished in terms of boundary conditions,
and so we classify the coordinates of a Dirichlet $p$-brane as follows:
\begin{eqnarray}
&&\mbox{} \left\{
\begin{array}{l}
\overline{A}_i~~(i=0,\ldots,p)~~:~~ \mbox{Neumann directions} \quad 
\partial_{\bf n}X^{\overline{A}}
\equiv \partial_{\bf n}X^{M}e_M^{\overline{A}}  = 0 
\\
\underline{A}_j~~(j=p+1,\ldots,10)~~:~~ \mbox{Dirichlet directions}
\quad \partial_{\bf t}X^{\underline{A}}
\equiv \partial_{\bf t}X^Me_M^{\underline{A}} = 0
\end{array}
\right.\,,
\end{eqnarray}
where we have introduced a normal vector $n^a$ to the boundary
$\partial\Sigma$, and the normal and tangential derivatives on the
boundary defined by
\begin{eqnarray}
\partial_{\bf n} \equiv n^a\partial_a\,, \quad \partial_{\bf t} \equiv
\epsilon^{ab}n_a\partial_b \quad (a,b=1,2)\,.
\end{eqnarray} 
Now we shall impose the boundary condition on the 
fermionic variable $\th$: 
\begin{eqnarray}
\label{p-} P^-\th |_{\partial \Sigma} = 0 \quad 
\mbox{or} \quad 
P^+\th |_{\partial \Sigma} = 0\,,
\end{eqnarray}
by using the projection operators $P^{\pm}$ defined by 
\begin{eqnarray}
P^{\pm} \equiv \frac{1}{2}\left( 1\pm s\,M^{10-p} \right)\,, \quad M^{10-p}
\equiv \Gamma^{\underline{A}_{p+1}} \Gamma^{\underline{A}_{p+2}} \cdots
\Gamma^{\underline{A}_{10}} \,.
\end{eqnarray}
We choose $s=1$ when both of $(+,-)$ are Neumann directions,
while $s=i$ when both of $(+,-)$ are Dirichlet directions.
Except for these cases, $P^\pm$ cannot be projection operators
because $(\Gamma^\pm)^2=0$.
We can express $\th$ as $\th = P^+\th$ on the boundary due to the
condition (\ref{p-}). The requirement that $P^{\pm}$ are projection
operators restrict the value $p$ to $p=1,2,5,6$ and 9.

From now on, we examine the $\kappa$-invariance of the action
(\ref{Lag}) on the pp-wave background (\ref{KG}).  Under the
$\kappa$-variation
\begin{eqnarray}
\delta_\kappa E^A=0~~~ \to ~~~ \delta_\kappa
X^M=i\thetab\Gamma^M\delta_\kappa\theta ~+{\cal O}(\theta^4)\,,
\label{kappa}
\end{eqnarray}
the action leads to a surface term only because non-surface terms vanish
for the supergravity solution.
The kinetic term
${\cal L}_0$
does not contribute to the
surface term.  The variation of ${\cal L}_0$ includes $\delta_\kappa
E_i^A=\partial_i(\delta_\kappa X^{\hat M})E_{\hat M}^A$ with $\hat
M=(M,\alpha)$, but the surface term vanishes because $\delta_\kappa
X^{\hat M}E_{\hat M}^A=\delta_\kappa E^A=0$.
On the other hand,
if we include $B^{(1)}$ in (\ref{first}), then
the studies become difficult.
It is because
the three-form contribution
on the boundary modifies the fermionic conditions.
The issue can be resolved by having additional degrees of freedom
at the boundary of the membrane,
and thus
we can consider
 the boundary surfaces with vanishing three-form
field strength\cite{dWPP:C}.\footnote{
We can couple a two-form field to the boundary surfaces
but we can make such a contribution vanish in the same way.}
It is thus enough to
investigate the surface term of the $\kappa$-variation of the
Wess-Zumino term $B^{(2)}$.

We first consider the Wess-Zumino terms without the terms including
$\mu$:
\begin{eqnarray}
S_{WZ}^{(\mu=0)} &=& \int \!\! d^3\sigma\,\epsilon^{ijk} \left[
\frac{i}{2}\bar{\th}\Gamma_{AB}\partial_i\th \cdot
\partial_jX^A\partial_k X^B +
\frac{1}{2}\bar{\th}\Gamma_{AB}\partial_i\th \cdot
\bar{\th}\Gamma^{A}\partial_j\th \cdot \partial_kX^B \right]\,,
\label{S:mu=0}
\end{eqnarray}
which leads to the well-known conditions in flat space.  The surface
terms come from the variation of variables with a derivative.  Under the
variation (\ref{kappa}), the action (\ref{S:mu=0}) leads to a surface
term up to and including fourth order in $\th$
\begin{eqnarray}
\del_{\kappa}S_{WZ}^{(\mu=0)} &=&
-\frac{1}{2}\int\!\!d\tau\int_{\partial\Sigma}\!\!d\xi\,
\Biggl[ \partial_{\bf t}
X^{\overline{B}} \left(
i\bar{\th}\Gamma_{\overline{A}\overline{B}}\del_{\kappa}\th \cdot
\dot{X}^{\overline{A}} + \bar{\th}\Gamma_{A\overline{B}}\dot{\th}\cdot
\bar{\th}\Gamma^A\del_{\kappa}\th +
\bar{\th}\Gamma_{A\overline{B}}\del_{\kappa}\th \cdot
\bar{\th}\Gamma^A\dot{\th} \right) \nn \\ && -
\dot{X}^{\overline{B}}\left( i\bar{\th}\Gamma_{\overline{A}
\overline{B}}\del_{\kappa}\th\cdot \partial_{\bf t}X^{\overline{A}} +
\bar{\th}\Gamma_{A\overline{B}}\partial_{\bf t}\th\cdot
\bar{\th}\Gamma^A\del_{\kappa}\th +
\bar{\th}\Gamma_{A\overline{B}}\del_{\kappa}\th\cdot \bar{\th}\Gamma^A
\partial_{\bf t}\th \right) \Biggr]\,,
\end{eqnarray}
where a dot on a variable means the world-volume time derivative
$\partial_\tau$ of the variable.
In order for the surface terms to vanish,
the following conditions must be satisfied
\begin{eqnarray}
&& \bar{\th}\Gamma_{\overline{AB}}\del_{\kappa}\th =
\bar{\th}\Gamma_{\underline{A}}\del_{\kappa}\th = 0\,.
\end{eqnarray}  
These conditions restrict the value $p$ to $p=1,5$ and 9.  Thus, we have
rederived the well-known result in flat space \cite{EMM,dWPP}.

In the case of pp-waves, there are the additional terms including the
parameter $\mu$ and so we have to take account of these terms.  It
is convenient to divide the Wess-Zumino terms including the parameter
$\mu$ into the following three parts:
\begin{eqnarray}
S_{WZ}^{(\mu)} &\equiv& 4!\cdot\frac{\mu}{2}
\int\!\!d^3\sigma\,\epsilon^{ijk}\partial_iX^C \partial_k X^B \Bigl[
i\bar{\th}\Gamma_{AB}T_C^{~+123}\th\cdot\partial_j X^A +
\bar{\th}\Gamma_{AB}T_C^{~+123}\th\cdot \bar{\th}\Gamma^A\partial_j\th
\nn \\ && - \bar{\th}\Gamma_{AB}\partial_j\th \cdot \bar{\th}\Gamma^A
T_C^{~+123}\th + \bar{\th}\Gamma_{AB}T_{C}^{~+123}\th \cdot
\bar{\th}\Gamma^AT_D^{~+123}\th \cdot \partial_j X^D \Bigr]\,,
\label{first}\\ S_{WZ}^{\rm spin} &\equiv& - \frac{1}{4}
\int\!\!d^3\sigma\, \Bigl[\bar{\th}\Gamma_{AB}\Gamma_{r-}\th
\bigl\{\partial_j X^A + \bar{\th}\Gamma^A\partial_j\th + 4!\mu
\bar{\th}\Gamma^A T_C^{~+123}\th \cdot \partial_jX^C \bigr\} \nn \\ && -
\bar{\th}\Gamma^A\Gamma_{r-}\th \bigl\{
\bar{\th}\Gamma_{AB}\partial_j\th +
4!\mu\bar{\th}\Gamma_{AB}T_{C}^{~+123}\th\cdot \partial_jX^{C} \bigr\}
\Bigr]\partial_iX^+ \partial_kX^B \partial^rG_{++}\,, \label{second}\\
S_{WZ}^{\mathcal{M}^2} &\equiv & \frac{i}{24}\int\!\!d^3\sigma\,\epsilon^{ijk}
\bar{\th}\Gamma_{AB}\mathcal{M}^2D_i\th\cdot\partial_jX^A\partial_kX^B\,,
\label{third}
\end{eqnarray}
up to and including fourth order in $\th$.  The first terms
(\ref{first}) include the parameter $\mu$ but do not have the spin
connection and $\mathcal{M}^2$ term.  The second terms (\ref{second})
include the spin connection without the $\mathcal{M}^2$ term, and the
only third term (\ref{third}) contains the $\mathcal{M}^2$ term.  

We analyze these terms in turn. 
Note that the boundary terms of the $\kappa$-variation
do not cancel out each other.
Therefore, all of these contributions have to be 
separately canceled out ubder appropriate additional boundary
conditions.
First, we consider the configuration of Dirichlet branes sitting at the
origin examining the Wess-Zumino term (\ref{first}).  As we will see
later, the other Wess-Zumino terms (\ref{second}) and (\ref{third}) do
not affect the result.  Secondly, we investigate the configuration of
Dirichlet branes sitting outside the origin.  In this case, the
Wess-Zumino term (\ref{second}) leads to additional conditions.

\subsection{Dirichlet Branes at the Origin}

Let us consider the Wess-Zumino terms (\ref{first}).  The variation of
(\ref{first}) under the $\kappa$-transformation leads to the following
surface terms:
\begin{eqnarray}
\del_{\kappa}S^{(\mu)}_{WZ} &=& -12\mu\int\!\!d\tau\int_{\partial\Sigma}
\!\!d\xi\, \Bigl[ \bigl\{ -
\bar{\th}\Gamma^{A}\del_{\kappa}\th\cdot\bar{\th}\Gamma_{\overline{C}
\overline{B}}
T_A^{~+123}\th +
\bar{\th}\Gamma_{A\overline{B}}T_{\overline{C}}^{~+123}\th\cdot
\bar{\th}\Gamma^A\del_{\kappa}\th \nn \\ && +
\bar{\th}\Gamma_{A\overline{B}}\del_{\kappa}\th\cdot \bar{\th}\Gamma^{A}
T_{\overline{C}}^{~+123}\th \bigr\}\dot{X}^{\overline{C}}\partial_{\bf
t}X^{\overline{B}} + \bigl\{ \bar{\th} \Gamma^A \del_{\kappa}\th \cdot
\bar{\th} \Gamma_{\overline{C}\overline{B}}T_{A}^{~+123}\th \nn \\ && -
\bar{\th}\Gamma_{A\overline{B}}T_{\overline{C}}^{~+123}\th\cdot
\bar{\th}\Gamma^A\del_{\kappa}\th -
\bar{\th}\Gamma_{A\overline{B}}\del_{\kappa}\th\cdot \bar{\th}\Gamma^{A}
T_{\overline{C}}^{~+123}\th \bigr\} \partial_{\bf
t}X^{\overline{C}}\dot{X}^{\overline{B}} \Bigr]\,.
\end{eqnarray} 
Since we have the conditions:
$\bar{\th}\Gamma^{\underline{A}}\del_{\kappa}\th =
\bar{\th}\Gamma_{\overline{AB}}\del_{\kappa}\th = 0$, we must impose
the additional constraints: 
\begin{eqnarray}
\bar{\th}\Gamma_{\overline{CB}}T_{\overline{A}}^{~+123}\th = 
\bar{\th}\Gamma^{\underline{A}}T_{\overline{C}}^{~+123}\th = 0\,.
\end{eqnarray}
These conditions restrict the $+,1,2,3$-directions, and so we have to
impose the boundary conditions for $+,1,2,3,$-directions as follows:
\begin{itemize}
\item One of $+,1,2,3$ is a Dirichlet direction and 
other three directions are Neumann directions. 
\end{itemize}
or 
\begin{itemize}
\item Three of $+,1,2,3$ are Dirichlet directions and 
the remaining one direction is a Neumann direction. 
\end{itemize}
That is, the directions of M5-branes, 9-branes and 1-branes are
restricted.  We shall classify the possible Dirichlet branes of an open
supermembrane on the pp-wave background below (we use indices
$I,J,K,\ldots,$ as the value in $1,2,3$, and $I',J',K',\ldots,$ as from 4
to 9).  

First, let us consider the configurations of 9-branes, which is
classified as follows:

\vspace*{0.2cm}
$\bullet$~~\underline{{\bf Classification of 9-branes}} (at the origin)
\quad ($\sharp D =1$ and $\sharp N = 10$)
  \begin{enumerate}
   \item D: one of the $1,2,3$-directions, N: otherwise.  $M^1 = \Gamma^{I}$
  \end{enumerate}
The above solution 1 with $I=3$ corresponds to the 9-brane solution used
in the study of a heterotic matrix model on a pp-wave \cite{M}, and so
our result gives a support for the work \cite{M}.

Now we shall consider the relationship of 9-branes to D8-branes in the
type IIA string theory. To do this, we move to the rotational frame
and then compactify a transverse direction on $S^1$ \cite{Michelson}.
By choosing the two directions from $4,\cdots,9$,
we obtain the type IIA string theory discussed in \cite{SY4,HS-S}, and
hence we can compare our result with the classification of D-branes in
this type IIA string theory.
We find that
9-branes of the solution 1 correspond
to D8-branes in type IIA string theory via the
$S^1$-compactification of transverse Neumann direction, and our result
agrees with the classification of D-branes in the type IIA string theory
\cite{SSY,HPS,HS-D,SY4}. 
 
Next, we shall classify the configuration of M5-branes preserving a half
of supersymmetries at the origin.

\vspace*{0.2cm}
$\bullet$~~\underline{{\bf Classification of M5-branes}} (at the origin) \quad 
($\sharp D =5$ and $\sharp N = 6$)
   \begin{enumerate}
    \item D: 1,2,3 and two of $4,\ldots,9$, 
N: + and $-$-directions. $M^{5} = \Gamma^{123I'J'}$ 
    \item D: one of $1,2,3$, and four of $4,\ldots,9$, 
N: + and $-$-directions and otherwise.  $M^{5} = \Gamma^{II'J'K'L'}$  
    \item D: +, $-$, two of $1,2,3$, and one of $4,\ldots,9$, N:
	  otherwise. 
$M^5 = \Gamma^{+-IJI'}$
  \item D: +, $-$, and three of $4,\ldots,9$, N: otherwise. 
$M^5 = \Gamma^{+-I'J'K'}$
   \end{enumerate}
The configurations of M5-branes correspond to those of D4-branes and
NS5-branes in type IIA string theory through the $S^1$-compactification
of Neumann and Dirichlet directions, respectively.
To compare with the known results in type IIA string theory,
we concentrate
on the cases 1 and 2 in which both $+$ and $-$-directions satisfy the
Neumann condition.

First, we consider D4-branes obtained from wrapped M5-branes.
As before, rotating two Neumann directions,
say 4 and 9,
and then compactifying one of the directions on $S^1$,
we obtain the D4-brane configuration
in which D4-branes are spanned along
$(+,-,4,b_1,b_2)$,
where $b=5,6,7,8$,
and $(+,-,I,J,4)$
for the cases 1 and 2 respectively,
after the appropriate relabelling of the coordinates. 
The projection operators for the boundary conditions
are constructed from
\begin{equation}
\label{D4}
\gamma^{123b_3b_4}
~~~~\mbox{and}~~~~
\gamma^{I5678}\,, \qquad 
\end{equation}
where $\gamma^r$'s are $SO(9)$ gamma matrices.
The resulting D4-brane configurations (\ref{D4})
are consistent with the classification of
Refs.\,\cite{SSY,HPS,HS-D}.

Next, we shall discuss NS5-branes obtained from unwrapped
M5-branes.
After rotating two Dirichlet directions,
and then compactifying one of the directions on $S^1$,
we obtain the configurations of
NS5-branes spanned along
$(+,-,b_1,b_2,b_3,b_4)$ and $(+,-,I,J,b_1,b_2)$
from the cases 1 and 2 respectively.
Thus, we have found that
the configurations of NS5-branes in the type IIA string theory are
also restricted in the case of pp-wave.

In addition to these configuration, we can derive various brane configurations
in other type IIA string theories on pp-waves,
which are not found before.
In particular, our configurations of the transverse M5-branes
obtained above might be related to those
discussed in \cite{transverse5}.

In the end, we consider the 1/2 BPS configurations of 1-branes at the
origin.

\vspace*{0.2cm}
$\bullet$~~\underline{{\bf Classification of 1-branes}} (at the origin) \quad  
($\sharp D =9$ and $\sharp N = 2$)
   \begin{enumerate}
    \item N: $+,-$, D: $1\ldots,9$. $M^{9}=\Gamma^{1\cdots 9}$ 
    \item N: one of $1,2,3,$ and one of $4,\ldots,9$, 
D: $+,-$ and otherwise.  $M^9 = \Gamma^{+-IJI'J'K'L'M'}$
   \end{enumerate}
Finally, we shall summarize the above result in Table\,\ref{cl:tab}. 

\begin{table}
 \begin{center}
  \begin{tabular}{|c|c|c|}
\hline
     & $M^{10-p}$ (N:+,$-$) &  $M^{10-p}$ (D:+, $-$)     \\
\hline\hline
 $p=9$ &  $\Gamma^{I}$ &   \\
$p=5$ &   $\Gamma^{123I'J'}$, $\Gamma^{II'J'K'L'}$ &  
$\Gamma^{+-IJI'}$, 
$\Gamma^{+-I'J'K'}$            \\
$p=1$ &  $\Gamma^{123456789}$ & $\Gamma^{+-IJI'J'K'L'M'}$      \\
\hline       
  \end{tabular}
\caption{Classification of 1/2 BPS Dirichlet branes sitting at the
  origin. \footnotesize The result for the case that $+$ and $-$
  directions satisfy the Neumann condition holds for the light-cone case
  at the origin.}
  \label{cl:tab}
 \end{center}
\end{table}

We can easily show that the Wess-Zumino term $S_{WZ}^{\mathcal{M}^2}$
including $\mathcal{M}^2$ does not change the above classification.  The
variation of $S_{WZ}^{\mathcal{M}^2}$ is given by
\begin{eqnarray}
\del_{\kappa}S_{WZ}^{\mathcal{M}^2} &=&
 \frac{i}{12}\int\!\!d\tau\int_{\partial\Sigma}\!\!
d\xi\, \bar{\th}\Gamma_{\overline{A}\overline{B}}\mathcal{M}^2\del_{\kappa}\th\cdot \partial_{\bf t}X^{\overline{A}}\dot{X}^{\overline{B}}\,, 
\end{eqnarray}
and hence we obtain the condition 
\begin{equation}
\bar{\th}\Gamma_{\overline{AB}}\mathcal{M}^2\del_{\kappa}\th = 0\,.
\end{equation}
By the use of the expression of $\mathcal{M}^2$, this condition can be
rewritten as
\begin{eqnarray}
&& 2\mu \bar{\th}\Gamma_{\overline{A}\overline{B}}T_{\overline{C}}^{~+123}\th\cdot
\bar{\th}\Gamma^{\overline{C}}\del_{\kappa}\th - \frac{\mu}{288}
\bar{\th}\Gamma_{\overline{A}\overline{B}}
\Gamma_{CD}\th\cdot\bar{\th}\Gamma^{CD+123}
\del_{\kappa}\th \nn \\
&& \hspace*{4cm}- \frac{1}{288}
\bar{\th}\Gamma_{\overline{A}\overline{B}}\Gamma_{EF}\th\cdot\bar{\th}\Gamma_{CD}\del_{\kappa}\th \cdot F^{EFCD} = 0\,.
\label{cond-re}
\end{eqnarray}
The condition (\ref{cond-re}) is satisfied for our configuration, as
we can easily check.  The remaining problem is to take account of the
effect of the spin connection. This issue will be discussed in the next
subsection.  The spin connection does not affect the classification
at the origin, as we will see later, so we have completely classified
the Dirichlet branes sitting at the origin. 

It should be remarked that we can rederive the result in the light-cone
formulation for the case that both light-cone directions satisfy the
Neumann condition.  The work \cite{SY1} considered only the Dirichlet
branes sitting outside the origin. However, we can consider those
sitting at the origin, and then the same result as in this paper can be
rederived. As a matter of fact, when we consider the Dirichlet branes
sitting outside the origin, we can obtain the same result as in the work
\cite{SY1}.

In this paper we have considered the $\kappa$-variation boundary terms
up to and including fourth order in $\theta$.  We expect that our
classification of Dirichlet branes is true even at the higher order in
$\theta$.  In fact, there are some arguments~\cite{BPZ} which support 
that this is the case.

\subsection{Classification of Dirichlet Branes outside the Origin}

Here we will include the contribution of the spin connection (i.e., the
effect of the order higher than $\mu$).  The variation of the Wess-Zumino
term $S_{WZ}^{\rm spin}$ with spin connections is given by
\begin{eqnarray}
\del_{\kappa}S_{WZ}^{\rm spin} &=& 
- \frac{1}{4}\int\!\!d\tau\int\!\!d_{\partial\Sigma}\xi\, \partial^rG_{++}
\Bigl[
\bar{\th}\Gamma_{AB}\Gamma_{r-}\th \cdot \bar{\th}\Gamma^+\del_{\kappa}\th
\cdot \dot{X}^{A}\partial_{\bf t}X^{B} \nn \\ 
&& -\bar{\th}\Gamma_{AB}\Gamma_{r-}\th\cdot \bar{\th}\Gamma^{A}\del_{\kappa}\th
\cdot \dot{X}^+\partial_{\bf t}X^{B} 
- \bar{\th}\Gamma_{AB}\del_{\kappa}\th\cdot \bar{\th}\Gamma^A\Gamma_{r-}\th
\cdot  \dot{X}^+\partial_{\bf t}X^{B} \nn \\
&& - \bar{\th}\Gamma_{AB}\Gamma_{r-}\th\cdot \bar{\th}\Gamma^+\del_{\kappa}\th
\cdot \partial_{\bf t}X^{A}\dot{X}^B 
+ \bar{\th}\Gamma_{AB}\Gamma_{r-}\th\cdot \bar{\th}\Gamma^A\del_{\kappa}\th
\cdot \partial_{\bf t}X^+ \dot{X}^B \nn \\
&& + \bar{\th}\Gamma_{AB}\del_{\kappa}\th\cdot \bar{\th}\Gamma^A\Gamma_{r-}\th
\cdot \partial_{\bf t}X^+\dot{X}^{B}
\Bigr] + \mathcal{O}(\th^6)\,,
\end{eqnarray} 
from which we can read off the additional conditions that surface terms
should vanish.  When the $+$-direction satisfies the Dirichlet
condition, the above surface terms all vanish, and hence we have no
further condition.  When we consider the case that $+$-direction
satisfies the Neumann condition, we have to impose the extra condition:
\begin{eqnarray}
\label{spin-cond}
\bar{\th}\Gamma_{\overline{A}\overline{B}}\Gamma_{\underline{r}-}\th \cdot 
\partial^{\underline{r}}G_{++} = 0\,. 
\end{eqnarray} 
It is a simple exercise to show that this condition holds for the
following two cases
\begin{eqnarray}
&& X^{\underline{r}}=0\,, \quad  \forall~\underline{r} \in \mbox{~Dirichlet
 directions}, \quad \\
&\mbox{or}& \quad 
\mbox{only $+$ and $-$ are Neumann directions}.
 \label{second-cond}
\end{eqnarray}
That is, the condition (\ref{spin-cond}) holds for all configurations
sitting at the origin. On the other hand, the Dirichlet $p$-branes with $p\neq 1$
outside the origin, i.e. $X^{\underline{r}}\neq 0$, does not satisfies 1/2 BPS condition if $+$-direction
satisfies the Neumann condition. 
The case (\ref{second-cond}) implies that the Dirichlet 1-branes
are 1/2 BPS objects even outside the origin.  Notably, the second
condition (\ref{second-cond}) is nothing but the result of the work
\cite{SY1}, and hence this result is consistent with the result of
Ref.\,\cite{SY1}. Finally, we shall summarize the result of the
classification outside the origin in Tab.\,\ref{out:tab}.  

\begin{table}
 \begin{center}
  \begin{tabular}{|c|c|c|}
\hline
     & $M^{10-p}$ (N:+,$-$) &   $M^{10-p}$ (D:+, $-$)     \\
\hline\hline
 $p=9$ &    &   \\
$p=5$ &    & 
$\Gamma^{+-IJI'}$, 
$\Gamma^{+-I'J'K'}$                            \\
$p=1$ & $\Gamma^{123456789}$ & $\Gamma^{+-IJI'J'K'L'M'}$      \\
\hline       
  \end{tabular}
\caption{Classification of 1/2 BPS Dirichlet branes sitting outside the
  origin. \footnotesize In the light-cone case there is a 1-brane only. 
This result is consistent with the work \cite{SY1}. }
  \label{out:tab}
 \end{center}
\end{table}

\section{Conclusion and Discussion}

In this paper we have discussed the boundary surfaces (Dirichlet branes)
of an open supermembrane on the pp-wave background. Our analysis is
based on the covariant formulation. Hence we could study a wider class
of configurations of Dirichlet branes than those in the light-cone
formulation where the light-cone directions must inevitably obey the
Neumann boundary condition.  We have classified the possible
configurations of Dirichlet branes both at and outside the origin.  In
particular, the 1/2 BPS configurations of M5-branes and 9-branes have
been found.  Some of the configurations are still 1/2 BPS outside the
origin.

It is also an interesting problem to look for less supersymmetric
configurations such as 1/4 BPS or 1/8 BPS Dirichlet branes of an open
supermembrane on the pp-wave background.
In our analysis, we considered branes without world-volume fluxes.
Turning on the world-volume fluxes is an interesting issue
\cite{SMT},
but left to the future investigation.

The classification of D-branes on the curved backgrounds is a very
interesting issue, and hence the classification of D-branes on the
pp-wave background is intensively studied by many authors.  It is an
interesting problem to study the Dirichlet branes of the supermembranes
on the $AdS_{4/7}\times S^{7/4}$ backgrounds or superstrings on the
$AdS_5\times S^5$.
We can possibly apply the similar method for such
theories.
It may be useful to examine the alternative Wess-Zumino terms
for superstrings on the $AdS_5\times S^5$
proposed in the works \cite{HaSa} (see also \cite{HKS}).
We will
study in this direction in the future \cite{SaYo2}.

Furthermore, it is interesting to examine the heterotic M-theory on the
pp-wave background, which is recently discussed in Ref.\,\cite{M}, from
the viewpoint of supermembrane theory.

\vspace{10mm}

\noindent 
{\bf\large Acknowledgements}

We would like to thank Katsuyuki Sugiyama for useful discussion on open
supermembranes at the first stage.

\appendix 

\vspace*{1cm}
\noindent 
{\large\bf  Appendix}
\section*{Notation and Convention}
In this place we will summarize miscellaneous notation and 
convention used in this paper. 

\subsection*{Notation of Coordinates}
For the supermembrane in the eleven-dimensional curved space-time, 
we use the following notation of supercoordinates for its superspace:
\begin{eqnarray}
(X^M,\th^{\al})\,, \qquad M = 
(+,-,\mu)\,,~~\mu = 1,\ldots, 9\,, \nn
\end{eqnarray}
and the background metric is expressed by $G_{MN}$. 
The coordinates in the Lorentz frame is denoted by 
\begin{eqnarray}
(X^A,\th^{\bar{\al}})\,, \qquad A=
(+,-,r)\,,~~r=1,\ldots,9\,, \nn
\end{eqnarray}
and its metric is described by $\eta_{AB} =
\mbox{diag}(-1,+1,\ldots,+1)$\,.   
The light-cone coordinates are defined by 
$
 X^{\pm} \equiv \frac{1}{\sqrt{2}}(X^0 \pm X^{10}) 
$\,.

The membrane world-volume is three-dimensional and its coordinates
are parameterized by $(\tau, \sigma^1, \sigma^2)$\,. 
The metric on the world-volume is represented by $g_{ij}$\,. 

\subsection*{$SO(10,1)$ Clifford Algebra}

We denote a 32-component Majorana spinor as $\th$, and 
the $SO(10,1)$ gamma matrices $\Gamma^A$'s satisfy the 
$SO(10,1)$ Clifford algebra 
\begin{eqnarray}
 & & \{\Gamma^A,\,\Gamma^B\} = 2\eta^{AB}\,, 
\quad  \{\Gamma^M,\,\Gamma^N \} = 2G^{MN}
\,, \quad \Gamma^A \equiv e^A_M\Gamma^M\,, 
\quad \Gamma^M \equiv  e^M_A\Gamma^A\,,
\nn 
\end{eqnarray}
where the light-cone component of the $SO(10,1)$ gamma matrices are 
\begin{eqnarray}
&&\Gamma^{\pm} \equiv \frac{1}{\sqrt{2}}\left(\Gamma^0 \pm \Gamma^{10}\right),
 \quad \{ \Gamma^+,\, \Gamma^- \} = -2 \mathbb I_{32}\,.
\nn 
\end{eqnarray}
We shall choose that $\Gamma^0$ is anti-hermite matrix and others 
are hermite matrices.  In this choice the relation 
$(\Gamma^A)^{\dagger} = \Gamma_0\Gamma^A\Gamma_0$ 
is satisfied. 
The charge conjugation of $\th$ 
is defined by 
\begin{equation}
 \th^{\s C} \equiv \mathcal{C}\bar{\th}^{\s T}\,, \nn 
\end{equation}
where $\bar{\th}$ is the Dirac conjugation of $\th$ and is defined by 
$\bar{\th} \equiv \th^{\dagger}\Gamma_0$. The charge conjugation matrix 
$\mathcal{C}$ satisfies the following relation:
\begin{eqnarray}
(\Gamma^A)^{\s T} = - \mathcal{C}^{-1}\Gamma^A \mathcal{C}\,, 
\quad \mathcal{C}^{\s T} = - \mathcal{C}\,. \nn
\end{eqnarray}
For an arbitrary Majorana spinor $\th$ satisfying the 
Majorana condition $\th^{\s C}=\th$, we can easily show the formula
\begin{equation}
\bar{\th} = - \th^{\s T}\mathcal{C}^{-1}\,. \nn 
\end{equation}
That is, the charge conjugation matrix $\mathcal{C}$ is defined by
$\mathcal{C}=\Gamma_0$ in this representation. The $\Gamma^A$'s are real
matrices (i.e., ($\Gamma^A)^{\ast} = \Gamma^A$). We also see that
$\Gamma^r$ and $\Gamma^{10}$ are symmetric and $\Gamma^{0}$ is
skewsymmetric.

\end{document}